\documentclass[twocolumn,showpacs,preprintnumbers,amsmath,amssymb,prb]{revtex4}

\usepackage{epsfig}
\newcommand{\expct}[2]{\left\langle #1 \right\rangle_{#2}}
\newcommand{\T}{\mathcal{T}}
\newcommand{\C}{\mathcal{C}}
\newcommand{\D}{\mathcal{D}}
\newcommand{\N}{\mathbb{N}}
\newcommand{\I}{\mathbb{I}}
\newcommand{\rhobar}{\bar{\rho}}

\begin{document}

\title{Nonlinear transport through a dynamic impurity in\\ a strongly
interacting one-dimensional electron gas}
\author{T.~L.~Schmidt and A.~Komnik}
\affiliation{Physikalisches Institut,
Albert--Ludwigs--Universit\"at Freiburg, D--79104 Freiburg,
Germany}
\date{\today}

\begin{abstract}
 We analyze the transport properties of a Luttinger liquid with an
 imbedded impurity of explicitly time-dependent strength. We employ a
 radiative boundary condition formalism to describe the coupling to the voltage
 sources. Assuming the impurity time dependence to be oscillatory
 we present a full analytic perturbative result in impurity strength
 for arbitrary interaction parameter calculated with help of Coulomb
 gas expansion (CGE). Moreover, a full analytic solution beyond the above
 restriction is possible for a special non-trivial interaction
 strength which has been achieved independently by full
 resummation of CGE series as well as via refermionization technique.
 The resulting nonlinear current-voltage characteristic turns out to be very rich due to
 the presence of the additional energy scale associated with the
 impurity oscillation frequency. In accordance with the
 previous studies we also find an enhancement of
 the linear conductance of the wire to values above the unitary
 limit $G_0 = 2 e^2/h$.
\end{abstract}


\maketitle

\section{Introduction}
Unlike their higher-dimensional counterparts the one-dimensional
metals are known to constitute a distinguished universality class
of so-called Luttinger liquids (LL).\cite{haldane} Probably the
most spectacular of their properties is the zero-bias anomaly
(ZBA) which reveals itself as a vanishing electronic density of
states at the Fermi edge. Among other things it leads to zero
conductance of an LL quantum wire in the low energy sector as long
as at least one static impurity is
present.\cite{kanefisher,furusakinagaosa} This has been
convincingly demonstrated in a number of recent experiments on
single-walled carbon nanotubes (SWNTs),\cite{bockrath1,yao} whose
electronic degrees of freedom had been shown to be adequately
described by the LL theory.\cite{sammlung1,sammlung3}

The way towards these spectacular findings was not smooth and has
been marked by a number of controversies, almost all of which seem
by now to be resolved. Probably the most prominent of them is the
linear conductance of the clean LL. According to early theoretical
studies it should be equal to $G = g G_0$, where $G_0 = 2 e^2/h$ is
the universal conductance quantum while
\begin{eqnarray}    \label{gdefinition}
g=1/\sqrt{1 + U_0/(\pi v_F)}
\end{eqnarray}
is the LL interaction parameter produced by the bare interaction
amplitude $U_0$.\cite{kanefisher,apelrice} However, the experimental
results favored the picture where $G=G_0$.\cite{tarucha} The
subsequent theoretical efforts could reveal the physics behind this
phenomenon. The clue to the problem's solution is the proper
inclusion of the voltage sources into the theory, which can either
be done by imposing boundary conditions on the particle densities,
or using the so-called $g(x)$-model where the interaction parameter
is supposed to extrapolate outside of the sample to the
non-interacting value $g=1$ in the particle reservoirs modeling the
voltage
sources.\cite{maslovstone,ponomarenko,safischulz,eggergrabert,dolcini05}
Both approaches yield matching results also for the impurity
scattering thereby surpassing the previously widely used \lq voltage
drop\rq\ approach, where the voltage is supposed to change abruptly
at the impurity site.

In view of future technological applications in, e.g., the
emerging field of nanoelectronics, the obvious way forward in
understanding the properties of such one-dimensional systems
appears to be the study of their transport properties in presence
of time-dependent perturbations. In general, these can be realized
in form of AC voltages or oscillating impurities. Such a situation
is quite interesting from the physical point of view as the
additional energy scale associated with the time evolution may
change the phase diagram of the system. Probably the simplest
question one can answer is whether the zero-bias anomaly survives
under such conditions. Thus far a number of different approaches
have been used and quite interesting predictions have been made.
\cite{linfisher,sharmachamon1,sharmachamon,andreevmishchenko,
quantumevaporation,feldmangefen,devillard,chengzhou,makogon,naon,guinea,schmeltzer}
One spectacular observation is the dynamic conductance enhancement
(DCE) reported by Feldman and Gefen.\cite{feldmangefen} Most of
the actual studies including the just mentioned one are using the
voltage drop approach, which requires much care as soon as it is
being used in connection with the perturbation expansion in
impurity strength. One attempt to improve and enhance the study
has been undertaken,\cite{chengzhou} where the authors have used
the $g(x)$-model. One is, however, still quite far from complete
understanding of the problem. For the major part because up to now
only perturbative calculations have been undertaken and no
analytic solutions for \emph{arbitrary} impurity strength,
voltages and temperature exist. We would like to close this gap.
For one thing we employ the more justified approach (from our
point of view) to take care of the applied voltage by means of the
radiative boundary condition formalism.\cite{eggergrabert} This
method in connection with the Coulomb gas expansion enables one to
construct an alternative full perturbative series in impurity
strength. One particular advantage of this series is that for a
special interaction parameter $g=1/2$ it can easily be summed up,
thereby yielding a first \emph{exact analytic} solution in the
full parameter range. In order to cross check these results we
also performed the same calculation in the refermionization
representation.

The structure of the paper is as follows: in Section~\ref{sec:LL}
we are going to present the necessary ingredients of the LL
picture and show how to incorporate voltage sources in the
presence of a dynamic impurity within the boundary condition
framework. The ensuing investigation in the path integral
formalism (Section~\ref{sec:PI}) will then allow us to derive a
perturbative result valid for small impurity strength and for all
interaction parameters $g<1/2$ (Section~\ref{sec:PT}).
Furthermore, in Section~\ref{sec:g12} we shall derive an exact
result in the special case $g=1/2$. This result will be confirmed
by an alternative calculation based on refermionization. A short
summary of results and discussion of predictions for experiments
concludes the paper.

\section{Luttinger liquid}\label{sec:LL}
The universality class of strongly interacting, one-dimensional
electron systems in the low energy sector is that of the so-called
Luttinger liquid. The corresponding generic electron-electron
interaction potential is extremely short ranged and is basically
given by $U(x-y) = U_0 \delta(x-y)$. This particular feature
facilitates a diagonalization of the model Hamiltonian for a clean
spinless system (realistic electrons with spin and other systems
are discussed in Conclusions) of infinite length in terms of
canonically conjugate bosonic fields $\theta(x)$ and $\partial_x
\phi(x)$, which describe the plasmon like excitations of the
system. Then (from now on we set $\hbar = e = k_B =1$)
\begin{equation}\label{H0}
    H_{LL} = \frac{v}{2} \int dx \left[ g (\partial_x \phi)^2 +
    g^{-1} (\partial_x \theta)^2 \right] \, ,
\end{equation}
where $g$ is given in Eq.~(\ref{gdefinition}). Values $g<1$
correspond to a repulsive interaction, whereas $g>1$ describes an
attractive potential and $g=1$ reproduces the non-interacting Fermi
gas case. The sound velocity $v = v_F/g$ is generated by
renormalization of the Fermi velocity $v_F$ of the bare
non-interacting system. In this bosonization representation the
ordinary fermion field operators are given by simple exponentials of
linear combinations of the fields $\theta(x)$ and $\phi(x)$.

The relation between the electron density and the phase fields is
even simpler. The electron density in the wire can be written as a
sum of contributions for left- and right-moving electrons,
$\rho(x) = \rho_L(x) + \rho_R(x)$ where $\rho_{L,R}(x) =
\psi^\dag_{L,R}(x) \psi_{L,R}(x)$, $\psi_{L,R}(x)$ being the field
operators of left/right moving electrons. Then, the phase field
$\theta(x)$ has a simple physical interpretation as it is related
to the electron density operator and the current operator $I(x,t)$
via
\begin{eqnarray}
    \rho(x,t) & = &  \rho_L + \rho_R
    = \frac{1}{\sqrt{\pi}} \partial_x \theta(x,t), \label{theta_rho} \\
    I(x,t)    & = &  v_F [\rho_L - \rho_R] = -\frac{1}{\sqrt{\pi}} \partial_t \theta(x,t).\label{theta_i}
\end{eqnarray}
An insertion of a local pointlike scatterer at $x=0$ results in
two different contributions to the Hamiltonian. The first one is
the forward scattering, proportional to $\psi^\dag_L(0) \psi_L(0)
+ \psi^\dag_R(0) \psi_R(0)$. In the case of a static impurity it
can be gauged away and does not affect the conductance in any way.
In the dynamic case its influence can be investigated using the
conventional equation of motion method.\cite{quantumevaporation}
From now on we concentrate on the local backward scattering
\begin{eqnarray}              \label{RS}
H_{sc} = \lambda(t) \left[ \psi^\dag_L(0) \psi_R(0) + \text{h.c.}
\right] \, ,
\end{eqnarray}
which is known to change the static transmission profoundly. Using
the bosonization identity one finds it to translate into
\begin{equation}\label{Hsc}
    H_{sc} = \lambda(t) \cos\left[2 \sqrt{\pi} \theta(x=0)\right].
\end{equation}
Here, $\lambda(t)$ is the impurity strength which we assume to
depend explicitly on time. Such a coupling can be realized in
experiments in many different ways. The best controlled way is
probably an STM tip placed in immediate vicinity of the LL. A
time-dependent voltage difference between the tip and the wire then
acts as an impurity of time-dependent strength. Another possibility
could be a setup where the wire is locally subjected to a laser
radiation of changing intensity. The third option is a locally
enhanced coupling to the internal phonon mode of the wire itself, as
long as the back action of the impurity onto the phonon degree of
freedom is negligible.

As outlined in the Introduction, the question of how to incorporate
an applied voltage $U$ has been subject to controversy. Early works
assumed a local drop in chemical potential at the impurity site but
the result was a wrong prediction for quantities as basic as the
linear conductance. It turned out that the reason for this failure
was that only a part $V$ of the applied voltage $U$ drops at the
impurity. In this picture, the voltage drop assumption, which has
been in use for quite a while, corresponds to $V=U$. It was shown
though, that this assumption is true only if the impurity energy
scale is sufficiently large, which renders perturbative approaches
in $\lambda$ based on the voltage drop assumption questionable.
There are different remedies to this shortcoming of which we favor
the radiative boundary conditions (RBC) approach.

For a static impurity this method leads to boundary conditions for
the phase field $\theta$ at the ends of the
wire.\cite{eggergrabert} It turns out that this technique can
easily be generalized to the case of a time-dependent impurity. If
we restrict our analysis to periodic impurity strength variations
and calculate time-averaged quantities, then all RBC equations are
very conveniently replaced by the ones averaged over a period $\T
= 2\pi/\Omega$ of the impurity oscillation. For the boundary
condition itself one then obtains
\begin{equation}\label{bc}
    \frac{1}{\T} \int_0^\T dt \left( \frac{1}{g^2} \partial_x \mp \frac{1}{v_F} \partial_t
    \right) \expct{\theta(x=\mp L/2)}{} = \pm
    \frac{U}{2\sqrt{\pi}v_F} \, ,
\end{equation}
which is, in fact, a weaker requirement then the original one.
Nevertheless, its implementation in order to compute the non-linear
current-voltage characteristic using (\ref{theta_i}) turns out to be
more cumbersome than the static calculation.

\section{Path integral description}\label{sec:PI}
In order to solve the impurity scattering problem in an LL using
the RBC approach, it turns out to be beneficial to describe the
system in terms of a real-time bosonic path integral. Translating
the Hamiltonian $H_{LL} + H_{sc}$ we find that the action
functional of the system consisting of the LL and the
time-dependent impurity reads
\begin{eqnarray}\label{S}
    S[\theta]
& = &
    \frac{1}{2g} \int_\C dt \int dx \left[ \frac{1}{v}
    (\partial_t \theta)^2 - v (\partial_x \theta)^2 \right] \nonumber \\
& - &
    \int_\C dt \lambda(t) \cos\left[ 2\sqrt{\pi}
\theta(x=0,t)\right].
\end{eqnarray}
The time integration has to be performed on the Keldysh contour
$\C$ as we are investigating nonequilibrium properties of the
system. The partition function is then given by $Z = \int {\cal D}
\theta e^{i S[\theta]}$ where the path integration runs over all
phase fields $\theta(x,t)$ compatible with the boundary conditions
(\ref{bc}). This can be achieved by decomposing $\theta(x,t)$ into
a homogeneous solution $\theta_h$ satisfying equilibrium boundary
conditions and a particular solution $\theta_p$ satisfying
(\ref{bc}). The latter reads
\begin{equation}\label{thetap}
    \theta_p(x,t) =
    -\frac{g^2 V}{2\sqrt{\pi} v_F}|x| - \frac{U-V}{2\sqrt{\pi}}t
\end{equation}
where $V$ is an arbitrary parameter. In the case of a static
impurity, it was shown that $V$ can be interpreted as the
four-terminal voltage, i.e. the voltage drop at the impurity site.
Generally, $V<U$, but for large impurity strength $V \rightarrow U$.
In the case of a dynamic impurity this interpretation must be
slightly altered. In analogy to the static impurity case, we fix the
parameter $V$ by the requirement
\begin{equation}\label{V}
    \frac{1}{\T} \int_0^\T dt
    \expct{\partial_t \theta_h(x,t)}{} = 0.
\end{equation}
The average current is then calculated by means of (\ref{theta_i}).
Our choice of $V$ ensures that the only contribution comes from
$\theta_h$ and one finds the time-averaged current
\begin{equation}\label{Iavg}
    \bar{I} = G_0(U-V)  \, .
\end{equation}
This result suggests the interpretation of $V$ as the average
four-terminal voltage and the current $I_{BS} = G_0 V$ as a
backscattering current.\cite{feldmangefen}

Inserting the minimizing action $\theta = \theta_p + \theta_h$ into
the action functional (\ref{S}), using (\ref{thetap}), one obtains
an action functional only depending on $\theta_h$,
\begin{eqnarray}\label{Sh}
    S[\theta_h]
& = &
    \frac{1}{2g} \int_\C dt \int dx \left[ \frac{1}{v} (\partial_t
    \theta_h)^2 - v (\partial_t \theta_h)^2 \right] \nonumber\\
& - &
    \frac{eV}{\sqrt{\pi}} \int_\C dt \theta_h(x=0,t) \\
& - &
    \int_\C dt \lambda(t) \cos\left[ 2\sqrt{\pi} \theta_h(x=0,t) -
    (U-V)t \right]. \nonumber
\end{eqnarray}
Having derived this effective action, we define the generating
functional
\begin{equation}\label{Z}
    Z[\eta] = \int \D\theta_h \exp
    \left\{ i S[\theta_h] + i \sqrt{\pi} \int_\C dt \eta(t)
    (\partial_t \theta_h) \right\}
\end{equation}
which enables us to rewrite (\ref{V}) in terms of a functional
derivative as
\begin{equation}\label{V2}
    \frac{1}{\T} \int_0^\T dt \frac{\delta Z[\eta]}{\delta
    \eta(t)}\bigg|_{\eta = 0} = 0.
\end{equation}
We assume for simplicity that the auxiliary field $\eta(t)$ is the
same on both branches of the Keldysh contour, $\eta(t_-) =
\eta(t_+)$. The task of calculating the current has thus been
reduced to the calculation of the partition function (\ref{Z}).

\begin{widetext}
With the exception of the additional time-dependent factor
$\lambda(t)$, the ensuing calculation can be performed in a similar
fashion as in the static case. The final result reads
\begin{equation}\label{Z_CGE}
    Z[\eta] = \exp\left\{ -\int dt dt' \eta(t) \ddot{C}(t-t')
    \eta(t') - i g V \int dt \eta(t) \right\}
     \left( 1 +
    \sum_{m=1}^\infty Z_m[\eta] \right)
\end{equation}
where
\begin{eqnarray}\label{Zm}
    Z_m[\eta]
& = &
    (i\lambda)^{2m} \int \D_{2m}t  \sum_{\{u_j\}'} \exp \left\{ \sum_{j>k} u_j u_k C(t_j-t_k) +
    \sum_j u_j \left[ \int dt \eta(t) \dot{C}(t-t_j) - i (U-V+gV) t_j \right] \right\}
     \nonumber \\
& \times &
    \mu(t_{2m}) \sin[\pi g \eta(t_{2m})] \prod_{j=1}^{2m-1}
    \mu(t_j) \sin\left\{ \pi g \left[ \eta(t_j) + \sum_{k=j+1}^{2m} u_k \right]
    \right\}.
\end{eqnarray}
\end{widetext}
Here, for convenience in the following perturbative expansion, we
assumed the time dependence to be of the form $\lambda(t) = \lambda
\mu(t)$ with $\max[\mu(t)]=1$. The system can be regarded as a
one-dimensional array of discrete charges $u_i = \pm 1$ located at
positions $t_i$ ($0<i<2m$) on the real time axis which interact via
the potential\cite{weiss95}
\begin{equation}\label{C}
    C(t) = 2 g \ln \left[ \frac{\beta\Delta}{\pi} \sinh\left( \pi
    |t|/\beta\right) \right] \, ,
\end{equation}
where $\Delta$ is the cut-off energy, i.e. the conduction band
width. In order to have a non-vanishing result, one must further
require the charge neutrality $\sum_j u_j = 0$ which is indicated by
the prime in the sum.  Finally, we introduced the time-ordered
integration measure
\begin{equation}
    \int \D_{n} t = \int_{-\infty}^\infty dt_n \int_{-\infty}^{t_n}
dt_{n-1} \cdots \int_{-\infty}^{t_2} t_1.
\end{equation}
We would like to stress that thus far no approximations have been
made, the results (\ref{Zm}) and (\ref{Z_CGE}) are \emph{exact}
for arbitrary $g$ and $\lambda(t)$.

\section{Perturbation theory in $\lambda$}\label{sec:PT}
Equation (\ref{Z_CGE}) is very well suited for a perturbative
expansion in $\lambda$. We limit ourselves to the leading order
which corresponds to $m=1$ and which is proportional to $\lambda^2$.
Using an oscillating impurity $\mu(t) = \cos(\Omega t)$, one can use
(\ref{V2}) and (\ref{Z_CGE}) and the self-consistency condition
(\ref{V}) reads
\begin{eqnarray}\label{V_sc}
    V
& = & \pi\lambda^2 \left(\frac{\pi}{\beta\Delta}\right)^2
    \sin(\pi g) \\
& \times & \int_0^\infty d\tau \frac{\cos(\Omega \tau)}{\sinh^{2g}(\pi \tau/\beta)}
    \sin[ (U-V+gV)\tau ]. \nonumber
\end{eqnarray}
For future convenience we introduce dimensionless quantities by
measuring all energies in units of $\beta=1/T$. Hence, defining $u =
\beta U$, $v = \beta V$, $\lambda' = \beta \lambda$, $\Delta' =
\beta \Delta$ and $\Omega' = \beta \Omega$, one obtains
\begin{equation}\label{V_sc2}
    v = \alpha \sin(\pi g) \int_0^\infty dx
\frac{\cos(\Omega'x)}{\sinh^{2g}(\pi x)} \sin[ux-(1-g)vx]
\end{equation}
where the prefactor was absorbed in the renormalized impurity
strength $\alpha = \pi \lambda'^2 (\pi/\Delta')^{2g}$. In the
following, we shall use this self-consistency equation to determine
the time-averaged differential conductance $G_d = d \bar{I}(U)/dU$
at $U=0$ as well as the time-averaged conductance $G=\bar{I}/U$ at
finite $U$.

By virtue of (\ref{Iavg}), the differential conductance is given by
\begin{equation}
   G_d = G_0[1-V'(U)] = G_0[1-v'(u)]
\end{equation}
as a function of the derivative $v'(u)$. At $u=0$, this function can
easily be calculated from (\ref{V_sc2}) by deriving both sides with
respect to $u$ and using $v(0) = 0$. One obtains
\begin{equation}             \label{diffcond}
    v'(0) = \frac{1}{\frac{1}{\kappa} + 1 - g}
\end{equation}
where
\begin{equation}
    \kappa(\Omega') = \alpha \sin(\pi g) \int_0^\infty dx
\frac{x\cos(\Omega'x)}{\sinh^{2g}(\pi x)}.
\end{equation}
This integral converges for strong interaction $0<g<1/2$ and can
be calculated analytically. It turns out that the zero bias
anomaly, which is always present in static systems for $T=0$,
vanishes as soon as the impurity acquires its own dynamics for
$\Omega \neq 0$. It appears that the finite oscillation frequency
plays the role of the effective temperature. In order to
investigate the interplay between the effects of finite $\Omega$
and $T$, the normalization to $1/\beta$ is very convenient.

\begin{figure}
    \centering
    \includegraphics{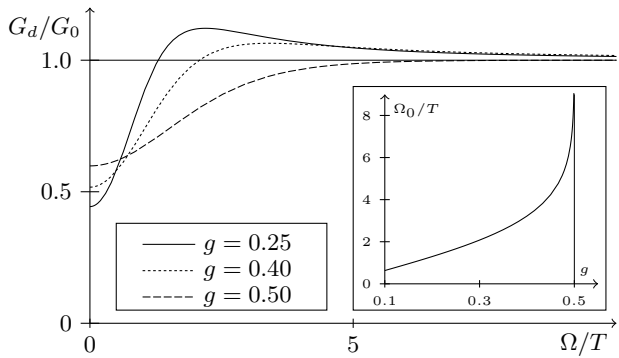}
    \caption{Perturbative differential conductance as function of impurity oscillation
    for various values of $g \leq 1/2$ for $\lambda'=0.8$ and $\Delta'=\pi$.
    $G_d$ exceeds $G_0$ at a critical frequency $\Omega_0$. {\em In the inset:}
    Dependence of the critical impurity frequency $\Omega_0$ on the interaction
    strength $g$}
    \label{fig_perturbDC}
\end{figure}

Using the last two equations one can analyze the differential
conductance. The resulting graph is shown in
Fig.~\ref{fig_perturbDC}. As expected, in the limit $\Omega
\rightarrow 0$ the remnants of the zero bias anomaly are more
pronounced for smaller $g$. The qualitative behavior of $G_d$
changes dramatically as soon as $g$ becomes smaller than $1/2$.
Then at some threshold value $\Omega_0$ the differential
conductance $G_d$ exceeds the unitarity limit $G_0$. This
dynamical conductance enhancement (DCE) is in accordance with
previous findings.\cite{feldmangefen,chengzhou} The actual
dependence of $\Omega_0$ on the interaction strength is quite
spectacular. While it is nearly linear for small $g$ (s. the inset
of Fig.~\ref{fig_perturbDC}), it diverges logarithmically towards
$g=1/2$. We find qualitatively the same behavior for the maximal
$G_d$ as a function of the interaction strength. By an explicit
calculation of the next order $\lambda$ contributions we made sure
that DCE is not a perturbation theory artefact.

\begin{figure}
    \centering
    \includegraphics{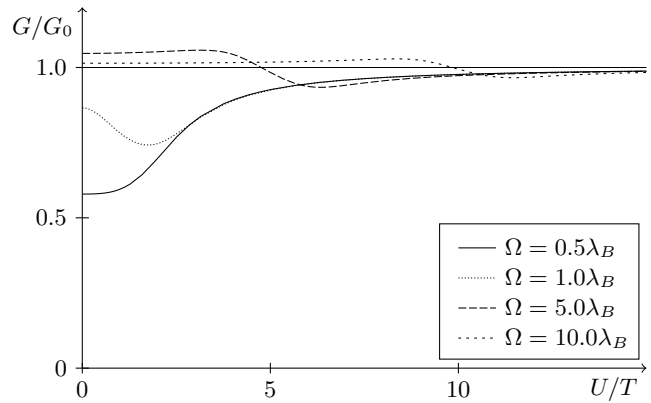}
    \caption{Perturbative linear conductance for $g=0.2$ as function of applied
    voltage $U$ for different values of $\Omega/\lambda_B$, where $\lambda_B = \pi \lambda^2/\Delta$.
    For small applied voltages the conductance exceeds $G_0$ for $g < 1/2$.}
    \label{fig_perturbIV}
\end{figure}

As a next step, we calculate the conductance for finite bias voltage
$U$. The integral in (\ref{V_sc2}) can be solved analytically for
$g<1/2$ and the resulting self-consistency equation can easily be
solved numerically, leading to a unique solution $v=v(u)$. By virtue
of (\ref{Iavg}) this leads to the current-voltage characteristic
shown in Fig.~\ref{fig_perturbIV}. The generic feature of all curves
is that for voltages higher then the oscillation frequency the
voltage dependence of the conductance approximately follows that of
the static system. This is not surprising as in this case the
excitations with energies of the order $U$ dominate the system's
transport properties rendering all other energy scales irrelevant.
Similar behavior has been found in the context of resonant tunneling
in LLs.\cite{a}

Again, for low voltages the conductance exceeds the conductance
quantum $G_0$ above $\Omega_0$. This behavior was predicted in the
voltage drop regime.\cite{feldmangefen} The authors also gave an
explanation of the effect: in the weak scattering regime $\lambda
\ll 1$, the system can be regarded as consisting of two chiral LLs
coupled by weak tunneling. Due to the $g$-dependent tunneling
density of states, scattering between left- and right-moving
particles may be strongly enhanced for $\Omega \approx U$. For
$g<1/2$ this can lead to a negative backscattering current. While
in the voltage drop regime this effect shows up whenever $\Omega >
U$, our results show that the occurrence is restricted to a
smaller parameter range. From Fig.~\ref{fig_perturbDC} and
Fig.~\ref{fig_perturbIV} we see that $G$ exceeds $G_0$ whenever
$\Omega > \max(\Omega_0, U)$.

\section{Exact solution at $g=1/2$}\label{sec:g12}
From the above results it became apparent that the interaction
strength $g=1/2$ is very special and in need of more thorough
investigation. Luckily, in this situation the summation of
(\ref{Z_CGE}) can be performed analytically and an exact result can
be derived. On the other hand, the corresponding Hamiltonian can be
rewritten in terms of some new fermionic fields (the so-called
refermionization procedure) thereby acquiring a quadratic shape.
This offers an alternative way to calculate the transport properties
of the system.

\subsection{Coulomb gas expansion resummation}
\label{sec:CGE}  The `collapsed-dipole-approximation'
\cite{weiss95} which after slight modifications continues to work
in the time-dependent case, becomes exact and using (\ref{V2}),
one obtains the self-consistency equation
\begin{eqnarray}\label{V3}
    V & = & \frac{2\pi \lambda_B}{\beta \T} \int_0^\T dt \mu(t)
    \int_0^\infty ds \frac{\sin[(U-V/2)s]}{\sinh[\pi s / \beta]} \\
& \times &
    \mu(t-s) \exp\{-\lambda_B \left[G(s) - G(0)\right] \} \nonumber
\end{eqnarray}
where $\lambda_B = \pi\lambda^2/\Delta$. In this equation, $G(s)$
is defined by $G'(s) = \mu^2(t-s)$ where no assumptions about the
actual form of $\mu(t)$ have been made. The result in the static
case is readily recovered as then $G(s) = s$.\cite{weiss95}

\begin{figure}
    \centering
    \includegraphics{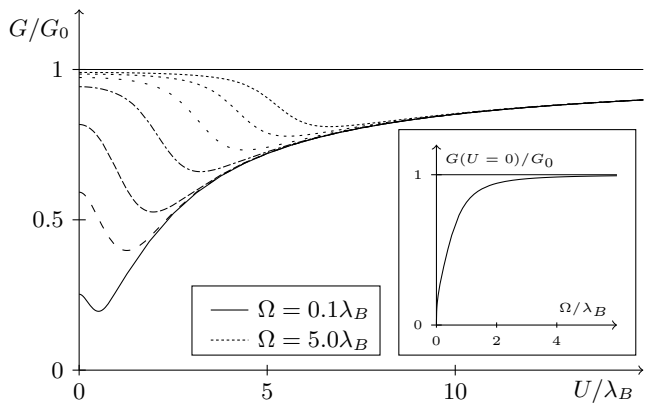}
    \caption{Conductance as a function of bias voltage for
    different impurity oscillation frequencies at zero
    temperature. {\em In the inset:} Weakening of the ZBA with
    increasing oscillation frequency.}
    \label{fig_CGE}
\end{figure}

It was also found that for the static case, all energies occurring
in (\ref{V3}) can be measured in units of the impurity energy scale
$\lambda_B$.\cite{eggergrabert} It turns out that this property is
maintained also the case of an oscillating impurity. We therefore
introduce the dimensionless quantities $v = V/\lambda_B$, $u =
U/\lambda_B$, $\vartheta = 1/(\beta \lambda_B)$ and $\epsilon =
\Omega/\lambda_B$. Now, the $t$-integration in (\ref{V3}) can be
performed and leads to
\begin{eqnarray}\label{V4}
    v
& = &
    \pi \vartheta
    \int_0^\infty dx
    \frac{\sin[(u-v/2)x ]}{\sinh\left(\pi \vartheta x\right)}
    e^{-x / 2} \\
& \times &
    \left\{ \cos(\epsilon x) I_0\left[- \frac{\sin(\epsilon x)}{2\epsilon}\right] +
    I_1\left[-\frac{\sin(\epsilon x)}{2\epsilon}\right]\right\}
    \nonumber
\end{eqnarray}
where $I_n(x)$ denotes the modified Bessel
function.\cite{abramowitz} The remaining convergent integral can
easily be evaluated numerically and the resulting self-consistency
equation can be solved and yields a unique voltage $v =
v_C(\epsilon, u, \vartheta)$. By virtue of (\ref{Iavg}), one obtains
the current-voltage characteristic shown in Fig.~\ref{fig_CGE}. As
the first prominent feature of this result one recognizes the
gradual disappearance of the zero bias anomaly. As in the
perturbative result, as soon as the applied voltage exceeds
$\Omega$, the current-voltage characteristic nearly reproduces the
static behavior.\cite{eggergrabert}

\subsection{Refermionization}\label{sec:RF}
The above results can be recovered by an alternative technique.
The starting point is the observation that the system Hamiltonian
$H = H_{LL} + H_{sc}$ can be rewritten in the following way
\begin{equation}\label{H_rf}
    H = -i v \int dx \Psi^\dag(x) \partial_x \Psi(x) + \lambda(t) a
    \left[ \Psi(0) - \Psi^\dag(0) \right]
\end{equation}
where $\Psi^\dag(x)$ and $\Psi(x)$ are new chiral fermionic fields
which depend on $\theta(x)$ and $\phi(x)$. The operator $a$ is an
auxiliary Majorana fermion satisfying $\{a, a\} = 2$ which does not
affect physical results. Insertion of such an object facilitates the
derivation of equations of motion for this
Hamiltonian.\cite{matveev95} Moreover, one can define a formal
particle density
\begin{equation}\label{rhotilde}
    \tilde{\rho}(x) := \langle : \Psi^\dag(x) \Psi(x) : \rangle
\end{equation}
where $\langle : \cdots : \rangle$ denotes normal ordering with
respect to the equilibrium $U=0$ state. This makes it possible to
express the boundary conditions (\ref{bc}) in terms of this new
density operator. As it is time-dependent, we take again the time
average over the oscillation period $\T$. Defining $\rhobar(x) =
(1/\T) \int_0^\T \tilde{\rho}(x,t)$, the boundary condition
(\ref{bc}) becomes\cite{eggergrabert}
\begin{equation}\label{bc_rf}
    5 \rhobar(-L/2) - 3 \rhobar(L/2) = \frac{U}{\pi v_F}.
\end{equation}
The Hamiltonian (\ref{H_rf}) can still be diagonalized by means of
the Heisenberg equations of motion. In the clean parts of the wire
($x\neq 0$) the solutions are plane waves with dispersion relation
$\omega = v k$. Due to the impurity potential, the function acquires
a discontinuity at $x=0$. Therefore, one makes the {\it ansatz}
\begin{equation}\label{psi_ansatz}
    \Psi(x,t) = \frac{1}{\sqrt{L}} \sum_k \exp\left[ i k (v t - x)
    \right] \left\{
              \begin{array}{ll}
                a_k, & \hbox{for $x<0$;} \\
                b_k, & \hbox{for $x>0$.}
              \end{array}
            \right.
\end{equation}
The anti-commutation relation then requires $\Psi(0,t) =
(1/2\sqrt{L}) \sum_k e^{ikvt} (a_k + b_k)$ at $x=0$. Inserting
(\ref{psi_ansatz}) into the Heisenberg equations of motions and
using the oscillating impurity strength $\lambda(t) = \lambda
\cos(\Omega t) =: \lambda \cos(v K t)$, one can derive a scattering
matrix equation for the fermions $a_k$ and $b_k$. Due to the
inelastic scattering, an operator $b_k$ will couple to operators
$a_{k'}$ and $a^\dag_{-k'}$ where $k' = k + 2 n K$ for all integers
$n \in \mathbb{Z}$. Therefore, it is convenient to introduce a
vector notation and to define
\begin{equation}\label{veca}
    {\bf a}(k) := \left(
                         \begin{array}{c}
                           \vdots \\
                           a_{k - 2K} \\
                           a_{k} \\
                           a_{k+ 2K} \\
                           \vdots
                         \end{array}
                       \right)
\end{equation}
and analogously for ${\bf b}(k)$. Moreover, we define the vector
${\bf a}^S(k)$ which is ${\bf a}(k)$ for $K \rightarrow -K$. The
scattering of $a_k$ fermions to $b_k$ fermions is then described
by the equation
\begin{eqnarray}\label{ab}
    {\bf b}^S(-k) & = &
    - \N(k) \left[ 2\N(k) - i \I \right]^{-1} \nonumber \\
& \times &
    \left[ i \N(k)^{-1} {\bf a}^S(-k)
    - 2 {\bf a}^\dag(k)\right]
\end{eqnarray}
where $\I_{mn} = \delta_{mn}$ is the unity matrix and $\N$ is
defined by ($m, n \in \mathbb{Z}$)
\begin{eqnarray}\label{N}
    \N(k)_{mn} & = & \frac{\lambda_B}{8 v} \bigg\{ \sum_{q=\pm} \frac{\delta_{m,n+q}}{k + (2n+q)K} \\
& + &
    \frac{\delta_{mn}}{k + (2n+1)K}
    + \frac{\delta_{mn}}{k +(2n-1)K} \bigg\} \nonumber
\end{eqnarray}
where we introduced the impurity energy scale $\lambda_B =
2\lambda^2/v$. This matrix is tridiagonal and can be inverted for
any finite dimension. Using a finite-dimensional matrix of size
$(n+1)\times(n+1)$ amounts to using $2 n \Omega$ as a cut-off
frequency. Therefore, although (\ref{ab}) is exact, numerical
results which base on a finite-dimensional matrix $\N(k)$ will be
more accurate for large oscillation frequencies $\Omega$.

The $a_k$ operators describe free fermions which constitute a Fermi
sea filled up to the chemical potential $eV$ which has to be
determined self-consistently from (\ref{bc_rf}). Then, the
time-dependent current can be calculated by $I(t) =
v_F\expct{\rho(x=0)}{}$.\cite{eggergrabert} For the time-averaged
current, we obtain formally the same result as for the static case,
\begin{equation}\label{I_rf}
    \bar{I} = \frac{v_F}{4} \sum_k \expct{(a^\dag_k + b^\dag_k)
(a_k + b_k)}{}.
\end{equation}
Going to a system of infinite length, the sum over $k$ can be
replaced by an integral and the self-consistency equation can be
solved numerically. The average current normalized with respect to
the unperturbed current $I_0 = G_0 U$ then reproduces the result in
Fig.~\ref{fig_CGE} obtained by CGE.

\section{Conclusion}
We have investigated the transport characteristics of a voltage
biased Luttinger liquid with an imbedded time-dependent impurity.
Using the radiative boundary condition formalism we derived an exact
analytic expansion for the nonlinear time-averaged current-voltage
characteristic in powers of the impurity strength. Due to the
presence of inelastic scattering processes brought about by the
dynamic behavior of the impurity, the conventional contact
resistance $h/2 e^2$ restriction can be overcome and higher
conductance emerges as long as the electronic correlations in the
system are strong enough. Furthermore, for the threshold value of
interaction strength we succeeded in a resummation of the
perturbation series and the transport properties of the system were
calculated \emph{exactly} in all regimes.

This effect of the dynamic conductance enhancement (DCE) relies
heavily on the special mathematical property of the backscattering
operator, namely on its scaling dimension $\alpha$ being equal to
$g$ in the spinless case. Moreover, it is then actually the local
operator with the smallest scaling dimension the theory can
support and the only one which can have $\alpha$ smaller than
$1/2$. This is the requirement to for observation of DCE. The
situation changes slightly when one is dealing with a spinful
system. In that case the corresponding Hamiltonian is constructed
out of two sets of conjugated fields describing the charge and
spin density fields $\theta_{c,s}$ and $\phi_{c,s}$. The electron
field operator is given by
\begin{eqnarray}
 \psi_{p=R/L, \sigma=\uparrow, \downarrow} \sim e^{ i \sqrt{\pi/2} \left[ p
  (\theta_c + \sigma \theta_s) -
 (\phi_c + \sigma \phi_s) \right] } \, .
\end{eqnarray}
The corresponding Hamilton operators for the participating fields
are given by two copies of (\ref{H0}) but with $g=1$ in the spin
channel.\cite{haldane} Hence the `physical' backscattering
operator as defined in (\ref{RS}) would only have the scaling
dimension $\alpha = (1+g)/2$ and therefore unable to generate DCE.
Nevertheless, there is an operator which has exactly the same
shape as (\ref{Hsc}) but contains the $\theta_c$ field only and
which has a scaling dimension $2g$. It corresponds to the more
complicated scattering of the kind $\psi_{R \uparrow}^\dag(0)
\psi_{L \uparrow}(0) \psi_{R \downarrow}^\dag(0) \psi_{L
\downarrow}(0)$. Not only would such a kind of scattering be
present in any spinful system with local scattering, it will also
be the most dominant one in the low energy regime because of its
small scaling dimension. Thus, for a spinful system with $g<1/4$
we expect our predictions to hold even \emph{quantitatively} as
long as $\Omega$ is not too high.

Probably the most spectacular realizations of LL physics are found
in the SWNTs. Contrary to the ordinary spinful LL the nanotube's
electronic degrees of freedom can be adequately described by a
four-channel LL. Three of them are free and only one of them is
genuinely interacting.\cite{sammlung1,sammlung3} Following the line
of reasoning from the last paragraph we shall find that the operator
with the smallest scaling dimension $\alpha = 4g$ can be constructed
out of 8 electron field operators and is given in Eq.~(6.10) of
[\onlinecite{epjb}]. The critical value for the interaction constant
is then $g=1/8$. As a rule, the correlation strengths measured in
the experiments vary between $0.1$ and $0.25$ (see
e.~g.~[\onlinecite{bockrath1}]). Therefore, we expect the DCE to be
visible in the SWNT-based setups, where the oscillating impurity is
either realized by an STM tip brought into the vicinity of the tube
or by a local coupling of the SWNT to a strong laser radiation.

\acknowledgments We would like to thank H.~Grabert and A.~O.~Gogolin
for many enlightening discussions. The authors are supported by the
DFG grant KO 2235/2.

\bibliography{dyn}

\end{document}